\documentstyle[aps,preprint]{revtex}
\begin{document}
\normalsize
\title{Theory for the Ultrafast Structural Response of Optically
Excited Small Clusters: Time Dependence of the Ionization
Potential}

\author{H. O. Jeschke, M. E. Garcia and K. H. Bennemann}

\address{Institut f\"ur Theoretische Physik der Freien
Universit\"at Berlin,
Arnimallee 14, 14195 Berlin, Germany.}

\maketitle

\widetext

\begin{abstract}
Combining an electronic theory with
molecular dynamics simulations we present results for the 
ultrafast structural changes in small clusters. 
We determine the time scale 
for the change from the linear to a triangular structure 
after the photodetachment process Ag$_3^- \rightarrow {\rm Ag}_3$. 
We show that the time-dependent change of the ionization potential
reflects in detail the internal degrees of freedom, in 
particular coherent and incoherent motion, and that it is sensitive 
to the initial temperature. We compare with experiment and point out
the general significance of our results.
\end{abstract}

\pacs{36.40.-c,31.70.Hq,32.80.Fb}


\newpage

The excitation of a cluster by a laser pulse induces time-dependent
changes of its electronic and atomic structure. These changes involve
bond formation and bond breaking. The understanding of the relaxation 
mechanisms induced by the excitation  is of general interest. 
In particular, it is of 
fundamental importance to study how the system approaches 
equilibrium in order to determine how the time-scales of the relaxation
processes can be controlled by varying the experimental conditions. 

Recently, a pump\&probe experiment has been performed on mass selected
Ag$_3^-$ clusters, which serves as an example for the investigation of 
the structural relaxation times\cite{nenepo}. 
The initially negatively charged clusters were 
neutralized through photodetachment by the pump pulse and after
a delay time $\Delta t$ ionized by the probe pulse in order to be detected. 
Due to the remarkable differences in the equilibrium geometries of 
the ground states of Ag$_3^-$, which is linear\cite{koutecky1},  
and Ag$_3$, which consists of an obtuse 
isosceles triangle\cite{koutecky2}, the ultrashort photodetachment
process puts the neutralized trimer in an extreme nonequilibrium
situation. As a consequence of that, a structural relaxation process
occurs. 
The experimental signal, consisting in the yield of
Ag$_3^+$ was measured as a function of  $\Delta t$ 
and the frequency of the probe laser pulse. For a frequency slightly above
the ionization potential (IP) of Ag$_3$ 
a sharp rise of the signal is observed at $\Delta t \simeq 750 {\rm
fs}$. After a maximum is reached, there is a saturation of the signal, 
which then remains constant for at least $100 {\rm ps}$, which 
is the longest time delay used in the experiment. 
New features appear for higher frequencies. Again the signal increases 
sharply, but after reaching a maximum it decreases to a constant value. 

A preliminary interpretation of these results uses 
the Franck-Condon-principle\cite{nenepo}. The first laser pulse 
creates a neutral linear silver trimer which
bends  and comes to a turning point near the equilateral
equilibrium geometry of the positive ion\cite{koutecky2}. After 
rebounding, the neutral trimer starts pseudo-rotating through its  
three equivalent obtuse isosceles equilibrium geometries. 
 This would explain the
saturation behaviour of the signal. However, this would mean that the
pseudo-rotations have an extremely long mean life, which seems improbable.
 Furthermore,  this model does not explain  why 
the signal changes as a function  of the frequency of the laser pulse.

In this paper we perform a theoretical analysis of the physics
underlying the ultrafast dynamics of Ag$_3$ clusters produced by
photodetachment. 
In particular, we analyze the time evolution of the ionization potential 
and the dependence of the dynamics on the initial temperature of the 
clusters. We show that the experimental results 
can be explained using a physical picture which can be generally applied 
to other ultrashort-time processes. In our calculations, 
we combine molecular dynamics (MD) simulations in the 
Born-Oppenheimer approximation with a microscopic theory to describe the
time-dependent electronic structure of the clusters.

In order to determine the potential energy surface (PES) needed for the MD
simulations, we start from a Hamiltonian of the form $H = H_{TB} + 1/2
\sum_{i \not= j} \phi({\bf r}_{ij})$, where the tight-binding part
$H_{TB}$ is
given by 
\begin{equation}
H_{TB} = \sum_{i,\alpha, \sigma} \varepsilon_{i \alpha} 
c_{i\alpha \sigma}^{+}c_{i\alpha \sigma}
  + \sum_{i \not= j, \sigma \atop \alpha,\beta} 
    V_{i\alpha j\beta} c_{i\alpha}^{+}c_{i\beta}. 
\end{equation}
Here, the operator $c_{i\alpha \sigma}^{+}$ ($c_{i\alpha \sigma}$) creates
(annihilates) an electron with spin $\sigma$ at the site $i$ and 
orbital $\alpha$ ($\alpha = 5s, 5p_{x}, 5p_{y}, 5p_{z}$). 
 $\varepsilon_{i \alpha}$ stands for the on-site energy, and 
 $V_{i\alpha j\beta}$ for the 
 hopping matrix elements. 
 For simplicity, and since the $5s$ electrons are expected to be rather 
 delocalized, we neglect the intraatomic Coulomb matrix elements.  
 $\phi({\bf r}_{ij})$ refers to the repulsive potential between the atomic
cores $i$ and $j$. For the distance dependence of the hopping elements
and the repulsive potential 
we use the functional form proposed in Ref.~\onlinecite{petifor}. 
By diagonalizing $H_{TB}$, (taking into account the angular
dependence of the hopping elements\cite{slater}), and summing over the 
occupied states, we calculate as a function of the atomic
coordinates the attractive parts of the electronic ground-state 
energies $E_{attr}^-$ and $E_{attr}^0$ of Ag$_3^-$ and Ag$_3^0$,
respectively. Then, by adding the repulsive part of $H$ we 
 obtain the PES, which we need to perform
the MD simulations. In order to determine the forces acting on the
 atoms we make use of 
 the Hellman-Feynman theorem. Thus, the $\alpha$-component of the force acting 
on atom~$i$ $F_{i \alpha} = - \partial E/\partial r_{i \alpha}$
is given by 
\begin{equation}
F_{i \alpha} = - \sum_{k \atop occ.} \langle k \mid 
        \frac{\partial H_{TB}}
       {\partial r_{i \alpha}} \mid k \rangle - \frac{1}{2} 
 \sum_{i \not= j} \frac{\partial \phi({\bf r}_{ij})}{\partial r_{i
\alpha}}. 
\end{equation}
Here the $\mid \! k \rangle$'s are the eigenstates of $H_{TB}$. 
The parameters of $H$ are determined in the following way. The on-site 
energies were obtained from atomic data\cite{moore}. 
We fit the parameters $V_{\alpha \beta}$ and the potential
$\phi({\bf r}_{ij})$ in order to reproduce 
 the equilibrium bond lengths of the Silver dimers 
Ag$_2^-$, Ag$_2$ and Ag$_2^+$ obtained by effective core 
potential-configuration interaction
calculations\cite{koutecky1,koutecky2}. We assumed the hopping elements
$V_{\alpha \beta}$ to fulfill Harrison's relations\cite{harrison}. 
The best fit was obtained by the following parameters: $V_{sp}=0.954\rm eV$,
$r_c = 4.33$\AA, $n_c = 2$, $m = 5.965$ and $A=0.605\rm eV$, where 
$r_c$ and $n_c$ refer to the cutoff radius and exponent,
whereas $m$ and $A$ stand for the exponent and strength of the
repulsive potential $\phi({\bf r}_{ij})$\cite{petifor}. 
Using these parameters we have calculated the vibrational frequencies of
the dimers, which compare reasonably well with the experimental 
values\cite{exptag2} and quantum-chemical 
calculations\cite{koutecky1,koutecky2}. 
Then, we determined the equilibrium geometries of the ground states of
the silver trimers Ag$_3^-$, 
Ag$_3$ and Ag$_3^+$ which again yielded excellent agreement with
the all-valence electron calculations of Refs.~\onlinecite{koutecky1} and
\onlinecite{koutecky2}. We have also determined a linear equilibrium geometry
for the excited state $^2\Sigma_g^+$ of Ag$_3$, which can be reached by
photodetachment of Ag$_3^-$ \cite{gantefoer}. However, in our calculations 
this state lies $1.2{\rm eV}$ higher than the ground state (in relatively good 
agreement with photodetachment experiments
\cite{gantefoer}) and therefore cannot be excited by the laser 
frequencies used in Ref.~\onlinecite{nenepo}. Therefore we 
consider in the following only the dynamics of the ground state of Ag$_3$.
Note, our hopping parameters are comparable to those of 
silver bulk\cite{papaconstantopulos}. This gives another justification
for the neglect of Coulomb interactions. 

The MD simulations are performed applying 
the Verlet algorithm in its velocity form. We used a time step of 
$\Delta t = 0.05fs$. This ensures an energy conservation up to 
$10^{-6} {\rm eV}$ after $10^{5}$ time steps. 
The equilibrium structures were obtained by performing simulated annealing.  
Starting with the equilibrium geometry of Ag$_3^-$, we generate an 
ensemble of approximately 1000 clusters characterized by the ensemble 
temperature T, defined as the time average of the kinetic energy for a long 
trajectory ($\sim 10^6$ time steps)\cite{jellinek}. 

In Fig.~1(a) we show the time dependence of the fraction of Ag$_3$ 
clusters $p(h\nu, t)$ with ionization potentials ${\rm IP}(t) \le 
h\nu$. We scale $h\nu$ with IP$_0$, which is the minimal ionization 
potential of Ag$_3$ in the equilateral equilibrium 
geometry of Ag$_3^+$. The quantity $p(h\nu, t)$
can be interpreted as the time-dependent probability for ionization
with a laser pulse of frequency $h\nu$ (i.e., half of this energy, if
the ionization is achieved in a two photon process as in 
Ref.~\onlinecite{nenepo}).
$p(h\nu, t)$ is proportional to the signal 
detected in the experiment. Note, slightly above the ionization threshold
($h\nu = 1.02{\rm IP}_0$), clusters can be ionized
only after a delay time $t_0(h\nu) \simeq 750 {\rm fs}$, at which $p(h\nu, t)$ 
begins to increase. $p(h\nu,t)$
displays the features observed in the experiment, namely first a sharp
increase, then a maximum and finally a decrease to a smaller value, which 
remains constant. There is, however, a narrow range of frequencies $h\nu$,
where the enhancement of the signal after the sharp increase is largest. 
For higher laser frequencies the maximum becomes 
broadened and the enhancement factor is smaller.

These results can be understood physically as follows. Upon 
photodetachment of a binding electron vibrational excitations occur,
in particular, those of the central
atom along the chain direction. 
Due to the shape of the PES the motion of the 
central atom dominates the
ultrashort time response over the first few hundred fs. Then, the slower 
thermally activated bending
motion comes into play and yields triangular bonded Ag$_3$. The
resultant bond formation is exothermic. The excess energy can in turn
cause bond breaking or in case of uniform energy
distribution also a regular vibrational mode like pseudo-rotations. 

In order to demonstrate the temperature dependence of the bond formation 
and bond breaking processes, we present in Fig.~1(b) results for the time 
dependent probability $p_{triang}(t)$ of finding triangular clusters
for different initial temperatures. $p_{triang}(t)$ is defined as the 
fraction of clusters having all angles between $45\deg$ and $90\deg$. 
Comparison of the behaviour of $p_{triang}(t)$ for 
${\rm T} = 317{\rm K}$ with the curves of Fig. 1(a) 
clearly shows that only these particular cluster geometries
contribute to the signal observed in the experiment\cite{nenepo}. 
Note that the onset of the abrupt increase of $p_{triang}(t)$ shows a
strong temperature dependence. For increasing initial 
temperature the clusters bend faster. 
Notice also the remarkable fact that at low temperature the maximum in
$p_{triang}(t)$ disappears. 
 
Fig.~2 shows contour plots for the time development of the distribution 
of the IP's of neutral silver trimers after photodetachment
at $t = 0$. The most prominent feature of these pictures
is the coherent oscillation of the IP
of all clusters, which continues for about $700{\rm fs}$ in the case
of clusters starting at ${\rm T} = 29{\rm K}$ and which is destroyed much
more quickly at the higher temperatures. These coherent oscillations of
the IP are due to the internal vibrations of the
linear chain before and during bending. High values of the
IP correspond to a rather symmetric geometry, 
while the lower values stand for the turning points of the 
oscillation. The second feature to be noted are the areas in the 
plot with IP's below approximately $1.07{\rm IP}_0$, 
that become
populated only after time delays of 1500, 700 and 300 fs respectively
and which represent those clusters that have bent to form geometries
close to the equilibrium geometry of Ag$_3$. For ${\rm T} = 1048{\rm K}$ we 
observed fragmentation of 31\% of the trimers (after $20{\rm ps}$), and 
in Fig.~2(c) we show only the distribution corresponding to the 
non-fragmented clusters.

With the help of Fig. 2 it is also possible to
explain the maximum of the IP as a result of the 
coherent vibrational motion of the clusters: while a large fraction 
of the clusters bend collectively,
they still continue their vibrational motion, which results in a 
pronounced minimum of the IP in the same way as for
the unbent trimers. 
This effect accounts for the dark areas in Fig.~2(b) and 2(c) for
${\rm IP}(t)/{\rm IP}_0$ below 1.07 approximately and is quantified by
integration over energies in the calculation of $p(h\nu, t)$ [Fig.~1(a)]. 
Since the clusters responsible for this effect have gained an average 
energy of $900{\rm K}$ while descending the PES towards 
the equilibrium structure of Ag$_3$, their coherent motion is quickly 
destroyed and this explains why no further maxima occur.
This is supported by the fact that the cluster ensembles 
starting out with temperatures of $29{\rm K}$ and $104{\rm K}$ do not show 
this maximum. Here, the coherent motion has already disappeared before the 
clusters bend toward the equilibrium geometry of Ag$_3$. 

Fig. 2 reveals that for longer times ($t > 2000{\rm fs}$) the clusters 
approach a time-independent distribution over the
IP's or equivalently over the Ag$_3$ geometries. 
These asymptotic distributions are plotted in Fig.~3(a).

In Fig. 3(b) we have plotted the quantity $p(h\nu, t)$ as in
Fig. 1(a), but for an initial temperature of ${\rm T} = 29{\rm K}$ and 
for a higher 
energy $h\nu = 1.17$IP$_0$. A new feature appears which consists in
sharp oscillations of the probability between 0 and 1 within the time
range $[0 \le t \le 700{\rm fs}]$. These oscillations result from  the 
dominant vibrational mode of the trimers for the first 
$700{\rm fs}$ approximately. This interesting effect was not observed in the
experiment\cite{nenepo}, because the laser energies used were
not high enough. In order to demonstrate that the excited state 
$^2\Sigma_g^+$ does not contribute to the signal observed in the experiment
of Ref.~\onlinecite{nenepo}, we have performed MD simulations
on the corresponding PES  and
confirmed its one dimensional dynamics. Its signal $p(h \nu, t)$ 
consist in 
   oscillations with no sharp increase like that 
observed in Ref.~\onlinecite{nenepo}. However, it is important to point out 
that this structure-less 
    signal 
  could 
  interfere with the bending or stretching dynamics of the  ground state 
 if the state $^2\Sigma_g^+$ were excited by the pump pulse.

The last important point to be investigated is the role of 
pseudo-rotations. In order to study these vibrational excitations we 
determine the autocorrelation function
\begin{equation}
G(\tau) = \frac{\sum_{k} q_x(t_k)q_x(t_k+\tau)}{\sum_{k}
q_x(t_k)q_x(t_k)}, 
\end{equation}
where $q_x$ is one of the normal coordinates of the triangular Ag$_3$
molecule\cite{martins}. If the pseudo-rotations had a long life
time, $G(\tau)$ would show an oscillatory behaviour. However, as shown
in the inset of Fig.~3(c), the correlations are strongly damped. This
becomes clear by analyzing the Fourier transform of $G(\tau)$. 
The life time of the pseudo-rotations, determined from the width of the 
peak at half maximum, is only slightly lower than the 
duration of one cycle. This shows that pseudo-rotations do not play an 
important role in the dynamics of the trimers. The high kinetic
energy of the atoms seems to prevent the occurrence of a regular
mode like pseudo-rotations. The saturation of the signal is rather a
statistical effect induced by the temperature.

In summary, by employing an electronic theory and MD simulations 
we have analyzed the femtosecond dynamics of
Ag$_3$ upon photodetachment. Our results show that an uncomplicated
theory which can easily be extended to deal with larger systems is 
able to account for all experimental observations. Most importantly, 
we find that the time scales for bond formation and breaking 
 is affected by temperature.  
Our results might also be of general interest regarding laser control of
chemical reactions. Fig.~3(a) suggests that temperature can be used to
design special distributions for the IP or for the cluster geometries 
 present in the initial ensemble.
Fig.~3(b) immediately suggests new experiments with a
probe laser pulse energy of $h\nu = 1.17{\rm IP}_0$.
Thus a direct measurement of the dominant vibration of the trimers seems
possible. 

This work was supported by the Deutsche Forschungsgemeinschaft.

\newpage

\begin{figure} 
\caption{
(a) Time dependence of the fraction of clusters $p(h\nu, t)$ with 
 ${\rm IP}(t)$ smaller than $h \nu$. The sharp 
increase  and the
overall time dependence of $p(h\nu, t)$ for increasing $h \nu$
should be compared with the experimental
results of Ref. 1. The initial temperature was ${\rm T} = 317 K$.
(b) Time dependence of the fraction of clusters $p_{triang}(t)$ having a
triangular structures  for different initial temperatures.
} 
\label{figure 1}
\end{figure}

\begin{figure}
\caption{ 
Distribution of photodetached clusters as a function of time 
and IP at different temperatures. Black regions 
indicate large number of clusters, whereas no clusters are present 
in the white parts. Note the progressive destruction of the coherent 
motion after photodetachment for increasing initial temperatures.
}
\label{figure 2}
\end{figure}

\begin{figure}
\caption{ (a) Average distribution of neutral silver clusters over
IP's for $2ps \le \Delta t \le 20ps$ after photodetachment.
(b) Calculated $p(h\nu, t)$ for a large value of $h \nu$. Note that  
 also the internal vibration
 of the chain before and during bending can be detected. 
(c) Spectral analysis of the pseudo-rotations.  
 Note that the life-time 
  is smaller than one pseudo-rotation period. }
\label{figure 3}
\end{figure}


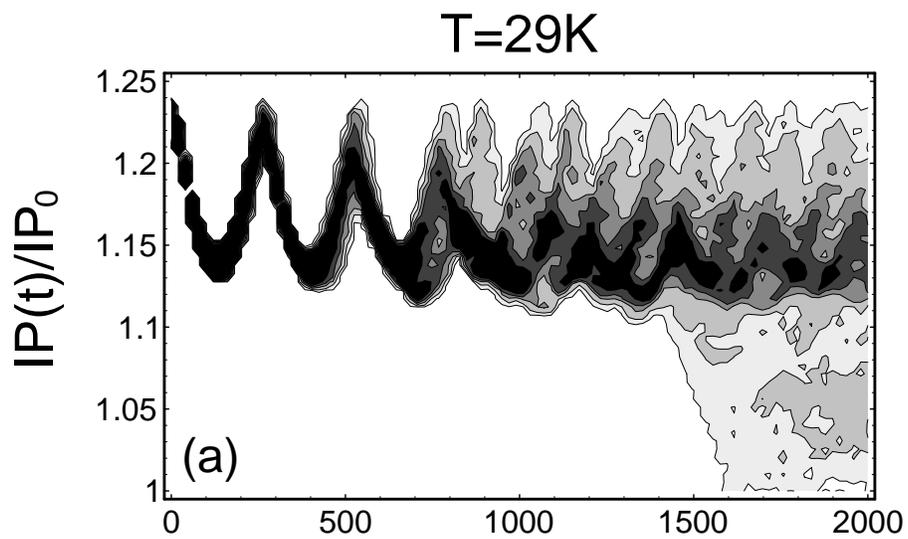
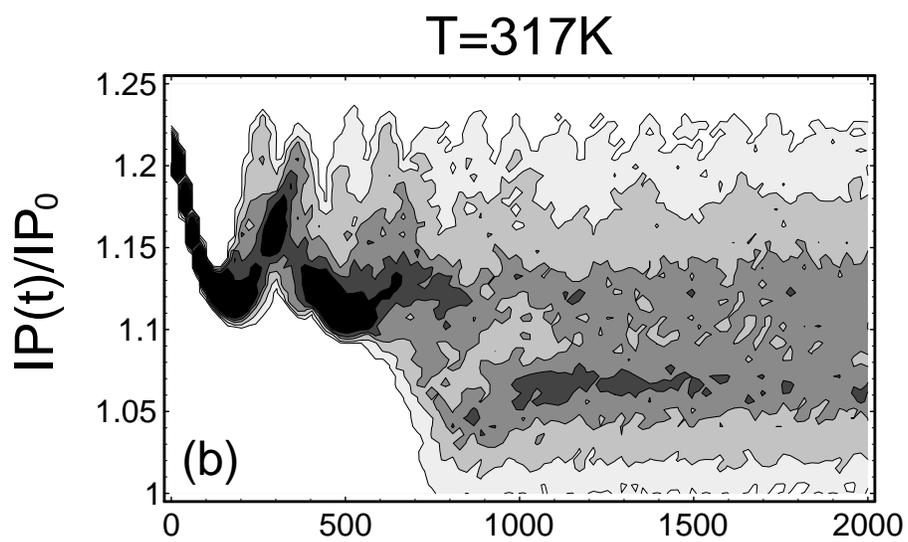
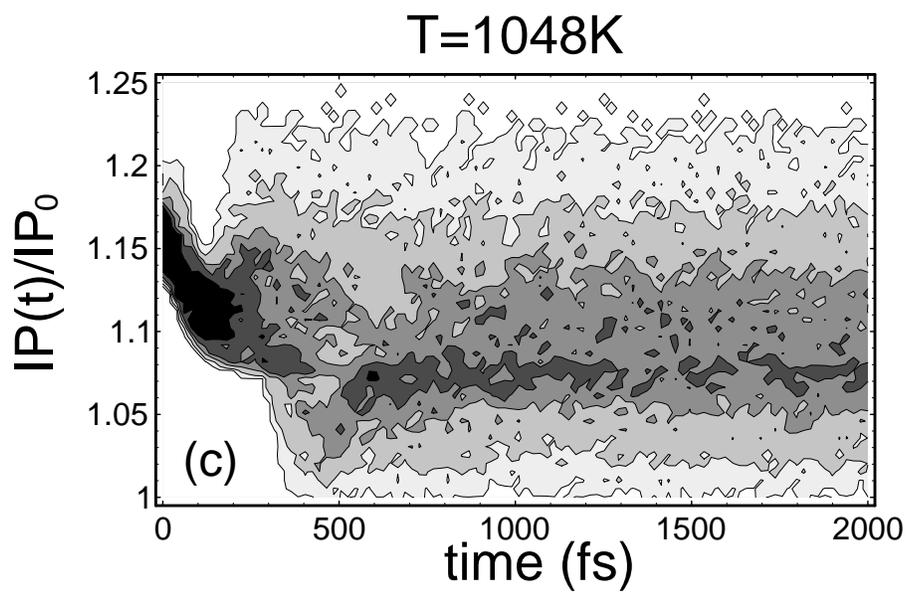
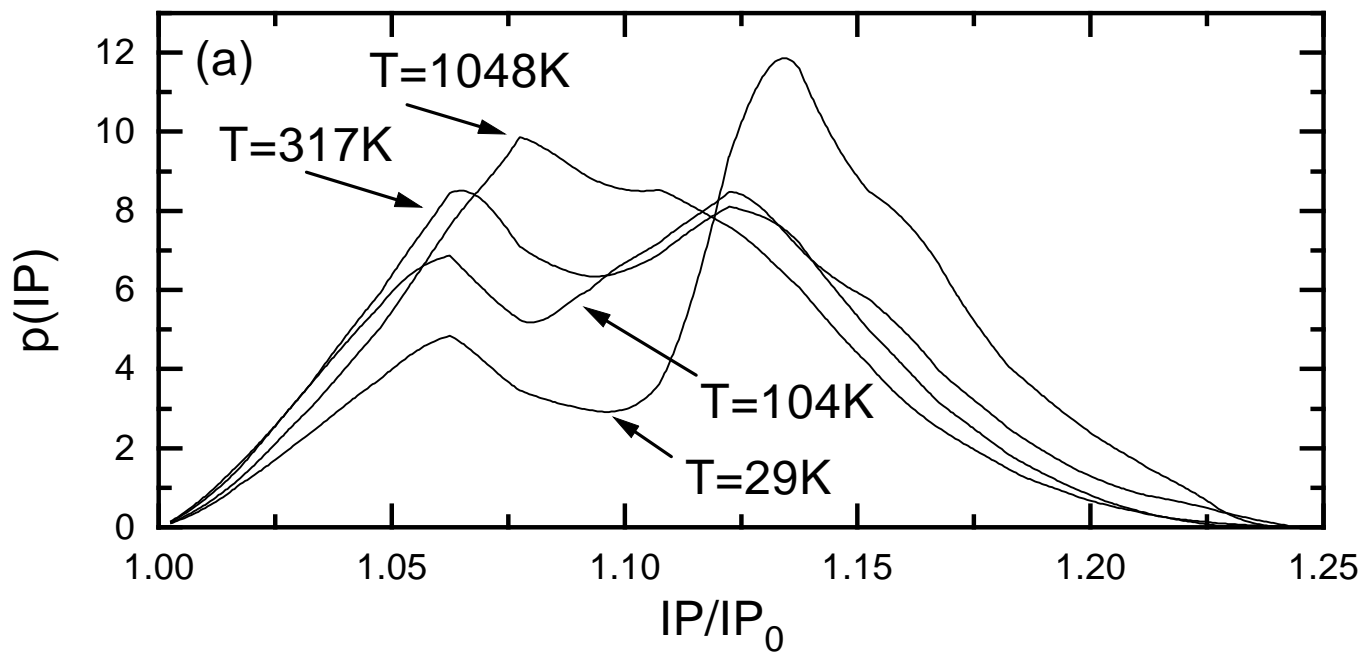
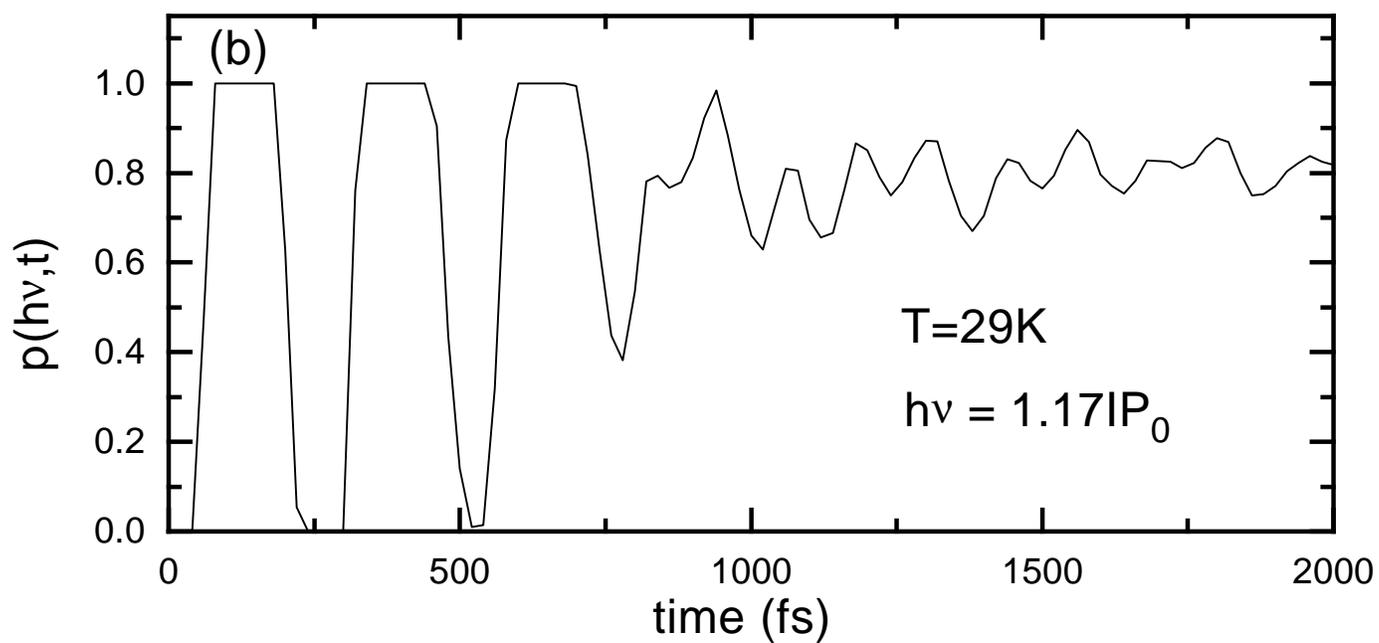
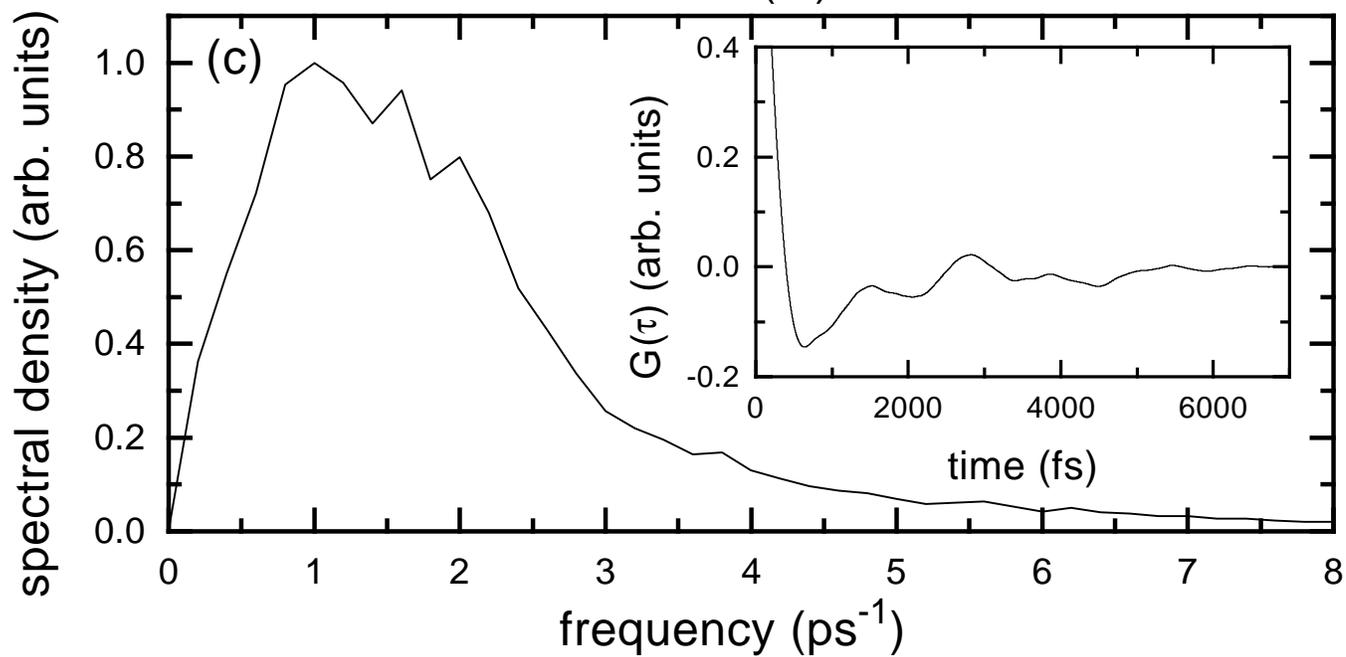
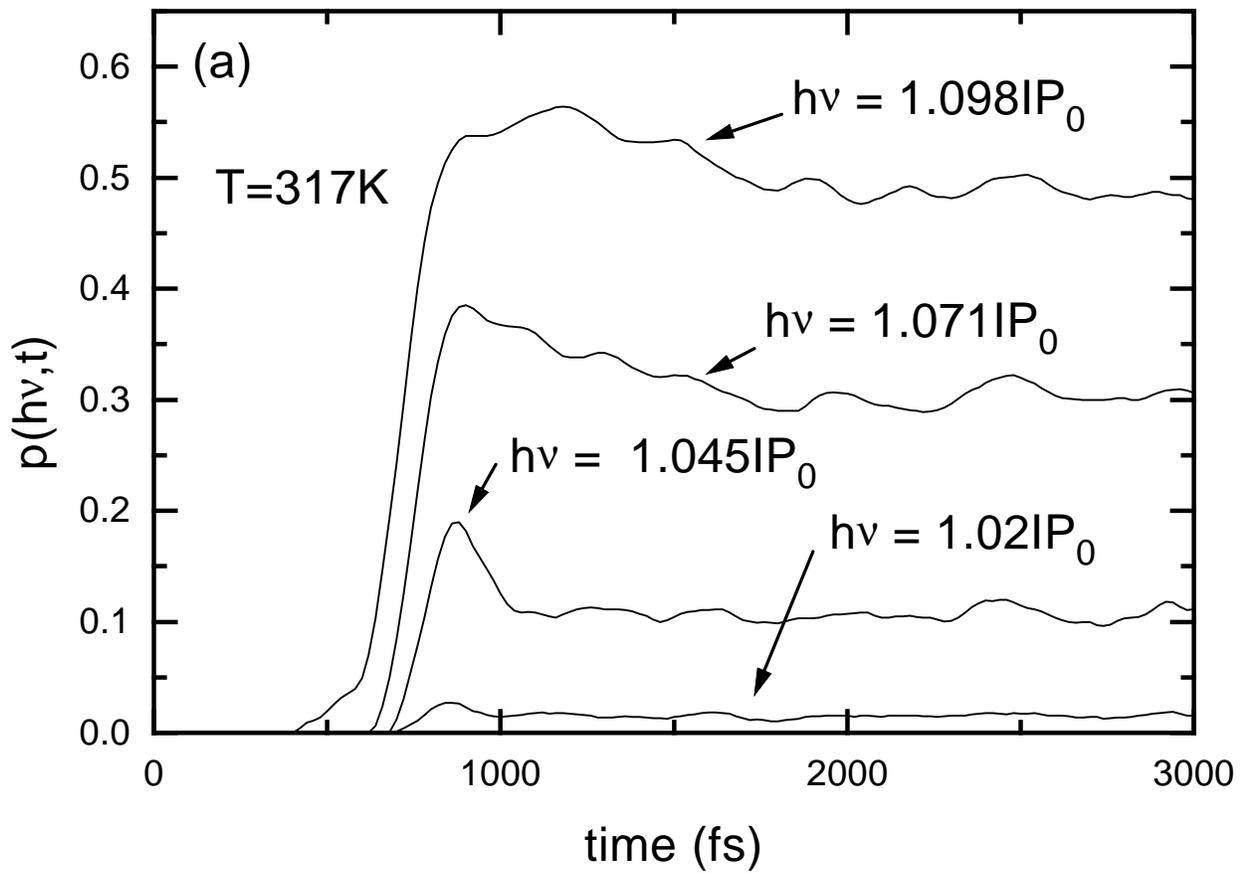
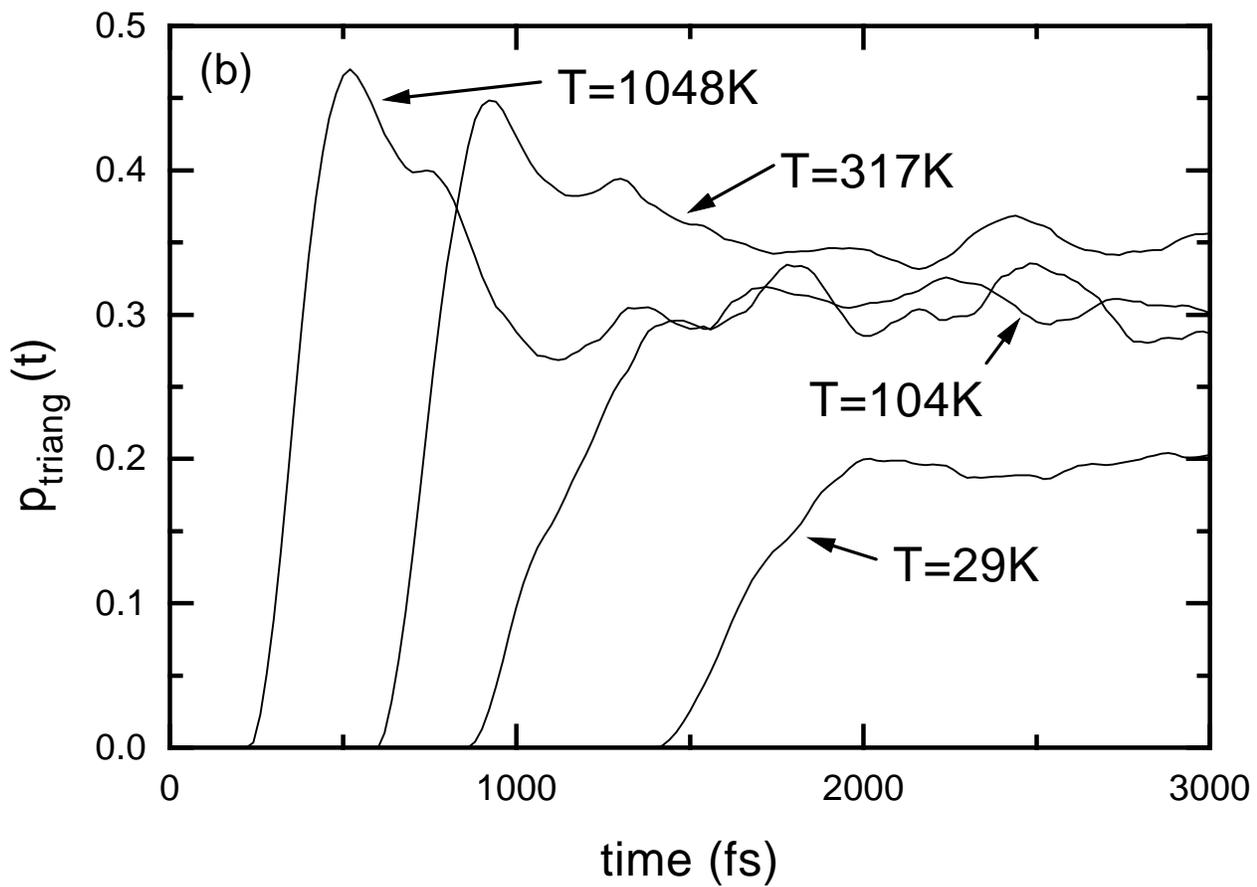
\end{document}